# Unveiling plasma energization and energy transport in the Earth's Magnetospheric System: the need for future coordinated multiscale observations

Alessandro Retinò[1]

Laboratoire de Physique des Plasmas, CNRS/Sorbonne Université/Université Paris Saclay/Observatoire de Paris/Ecole Polytechnique/ Institut Polytechnique de Paris, Paris, France
Email: alessandro.retino@lpp.polytechnique.fr
Phone: +33169335929

Co-authors:

L. Kepko[2], H. Kucharek[3], M.F. Marcucci[4], R. Nakamura[5], T. Amano[6], V. Angelopoulos[7], S. D. Bale[8], D. Caprioli[9], P. Cassak[10], A. Chasapis[11], L.-J. Chen[2], L. Dai[12], M. W. Dunlop[13], C. Forsyth[14], H. Fu[15], A. Galvin[3], O. Le Contel[1], M. Yamauchi[16], L. Kistler[3], Y. Khotyaintsev[17], K. Klein[18], I. R. Mann[19], W. Matthaeus[20], K. Mouikis[3], K. Nykyri[21], M. Palmroth[22], F. Plaschke[23], Y. Saito[24], J. Soucek[25], H. Spence[26], D. L. Turner[27], A. Vaivads[28], F. Valentini[29]

[1]Laboratoire de Physique des Plasmas, CNRS/Sorbonne Université/Université Paris Saclay/Observatoire de Paris/Ecole Polytechnique/ Institut Polytechnique de Paris, Paris, France; [2] NASA Goddard Space Flight Center, Greenbelt, MD, USA; [3]University of New Hampshire, Durham NH, USA; [4]INAF-Istituto di Astrofisica e Planetologia Spaziali, Rome, Italy; [5]Space Research Institute, Austrian Academy of Sciences, Graz, Austria; [6]School of Science, University of Tokyo, Japan; [7]University of California Los Angeles, CA USA; [8]University of California Berkeley, CA USA; [9]University of Chicago, USA; [10]West Virginia University, USA; [11]Laboratory for Atmospheric and Space Physics, Boulder, USA; [12]National Space Science Center, China; [13]RAL Space, STFC, UK; [14]UCL Mullard Space Science Laboratory, UK; [15]Beihang University, China; [16]Swedish Institute of Space Physics, Kiruna, Sweden; [17]Swedish Institute of Space Physics, Uppsala, Sweden; [18]University of Arizona, USA; [19]University of Alberta, Canada; [20]University of Delaware, Newark DE USA; [21]Embry-Riddle Aeronautical University, USA; [22]University of Helsinki, Helsinki, Finland; [23]IGEP, TU Braunschweig, Germany; [24]ISAS-JAXA, Japan; [25]IAP, Czech Academy of Sciences, Prague, Czechia; [26]Institute for the Study of Earth, Oceans, and Space, USA; [26]Johns Hopkins Applied Physics Laboratory, USA; [28]KTH, Stokcholm, Sweden; [29]Università della Calabria, Rende, Italy.

**Synopsis.**

Energetic plasma is everywhere in the Universe: planetary and exoplanetary magnetospheres, stellar coronae, supernova remnant shocks, accretion disks, and astrophysical jets. The terrestrial Magnetospheric System is a key case where direct measures of plasma energization and energy transport can be made in situ at high resolution, allowing quantitative understanding of the underlying plasma physics. Despite the large amount of available observations, including multi-point observations from Cluster, THEMIS and MMS, we still do not fully understand how plasma energization and energy transport work. This is essential for understanding how our planet works, e.g., how charged particles escape from our ionosphere/atmosphere, how solar-wind energy enters the boundaries of the magnetosphere, and couples with the ionosphere/atmosphere. Key physical processes driving much plasma energization and energy transport occur where plasma on fluid scales couple to the smaller ion kinetic scales and involve processes occurring in non-planar and non-stationary structures. These scales ($\lesssim$ 1 RE) are strongly related to the larger mesoscales (~ several RE) at which large-scale plasma energization and energy transport structures, e.g. large-scale plasma flows, form. All these scales and processes need to be resolved experimentally, however existing multi-point in situ observations do not have a sufficient number of measurement points. New multiscale observations simultaneously covering scales from mesoscales to ion kinetic scales are needed to make the next step in our understanding of plasma energization and energy transport and, more generally, of the structure and dynamics of the terrestrial Magnetospheric System. The implementation of these observations requires a strong international collaboration in the coming years between the major space agencies. The Plasma Observatory is a mission concept tailored to resolve scale coupling and non-planarity/non-stationarity in plasma energization and energy transport at fluid and ion scales. It targets the two ESA-led Medium Mission themes "Magnetospheric Systems" and "Plasma Cross-scale Coupling" outlined in the final recommendations of the ESA Voyage 2050 report and is currently under evaluation as a candidate for the ESA M7 mission, to be launched around 2037. MagCon (Magnetospheric Constellation) is a mission concept being studied by NASA aiming at studying the flow of mass, momentum, and energy through the Earth's magnetosphere at mesoscales. Coordination between Plasma Observatory and MagCon missions, together with possible additional synergies with multi-scale missions from other international agencies, would allow us for the first time to simultaneously cover from mesoscales to ion kinetic scales leading to a paradigm shift in the understanding of the Earth's Magnetospheric System.

**Introduction**

Examples of collisionless systems where strong and spectacular particle energization and energy transport events occur are stellar coronae and winds, planetary magnetospheres, supernova remnant shocks, accretion disks, astrophysical jets, and the large-scale intergalactic plasma permeating the cosmic web (Retinò 2021). A complete comprehension of particle energization and energy transport mechanisms in plasmas is far from being achieved. Understanding the physics behind these mechanisms is a compelling science problem of major importance for the space, solar, and laboratory plasma communities.

A cross-disciplinary synergy between the above communities is crucial to advance our understanding of particle energization and energy transport (Verscharen 2021). Remote solar observations have considerably increased their resolution in the last decade owing to data from Soho, Hinode, SDO, and more recent Parker Solar Probe and Solar Orbiter missions. However, they are still not adequate to resolve the full physics of particle energization and energy transport, in particular at small plasma scales. Despite the recent diagnostic improvements in laboratory plasmas, see e.g., MRX, TORPEX, FLARE, BaPSF, and NIF experiments, the laboratory setup often imposes severe limiting factors in terms of boundary conditions (Fasoli 2013; Fiuza 2020; Ji 2018; Yamada 1997). Due to the inherent complexity of the underlying physics, understanding particle energization and energy transport mechanisms in depth from an experimental point of view requires direct measurements of plasma and electromagnetic fields. This is currently possible only in the Heliosphere.

**The Earth's magnetospheric System**

The Earth's Magnetospheric System is the complex, dynamic region forming in near-Earth space upon the interaction of the solar wind with the Earth's magnetic field, see Figure 1. Intense, often explosive, plasma energization and energy transport processes occur therein. The incoming cold solar wind plasma gets actively energized up to millions of times when interacting with the near-Earth plasma environment, thus ultimately driving transport of vast amounts of energy into the Earth's Magnetospheric System. The latter represents the best natural environment in the Heliosphere to experimentally study the physics of plasma energization and energy transport, since refined in situ measurements of both electromagnetic fields and particle distribution functions can be made at multiple points and transmitted to the ground in large volumes at high cadence.

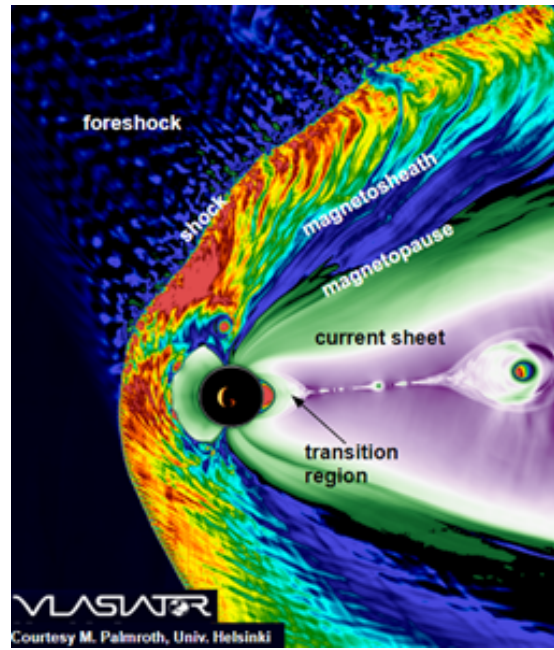

*Figure 1. The Magnetospheric System. Key regions for energy transport and particle energization are the bow shock, the magnetosheath, the magnetopause, the magnetotail current sheet, the transition region and the inner magnetosphere.*

These measurements are of pivotal importance to study key plasma energization and energy transport processes such as shocks, magnetic reconnection, turbulence and waves, plasma jets and instabilities. This allows us to ultimately understand how our planet works, including the very important aspect of space weather. These measurements also have a paramount impact on unveiling the dynamics and the energetics of a number of solar and astrophysical plasma environments with similar physical properties, as well as improving the comprehension of laboratory and fusion plasmas.

**Multi-scale plasma energization and energy transport**

The fundamental physical processes governing plasma energization and energy transport operate across multiple scales ranging from large mesoscales to fluid and then ion and electron kinetic scales (Retinò 2021; Schwartz 2009; Nykyri 2021). Fluid and ion scales are the scales at which the largest amount of electromagnetic energy is converted into energization of charged particles. Addressing fluid and ion scales and their coupling is hence a major endeavor for space, solar and astrophysical plasmas. Key examples of plasma energization and energy transport processes at ion scales and above are shocks, magnetic reconnection, turbulence and waves (e.g. Kelvin-Helmholtz), plasma jets and the combination of these processes. Numerical simulations also show that plasma energization and energy transport often occur in 3D structures which are very often nonplanar and nonstationary (Pezzi 2020; Pisokas 2018; Servidio 2014). Resolving the coupling between fluid and ion scales, as well as nonplanarity and nonstationarity, is therefore crucial. In situ measurements of electromagnetic fields and particle distributions at multiple points and covering simultaneously ion (≲1000 km) and fluid (between ~1000 km and ~1 RE) scales are thus required. These scales are strongly related to the larger mesoscales (~ several RE) at which large-scale plasma energization and energy transport structures, e.g. large-scale plasma flows, form. All these scales and processes need to be resolved experimentally, however existing multi-point in situ observations do not have a sufficient number of observation points. The Earth's Magnetospheric System is the best space laboratory in the Heliosphere where these observations can be performed.

**State of the art: comparison with existing and upcoming multi-point in situ measurements**

Over the last two decades, ESA/[Cluster](Cluster) and NASA/[MMS](MMS) four-point constellations, as well as the large-scale multipoint mission NASA/[THEMIS](THEMIS), have tremendously improved our understanding of plasma processes at individual scales compared to earlier single-point measurements. However, such missions have at most 4 points at a single scale and were not designed to study scale coupling. Reaching closure on plasma energization and energy transport physics requires understanding how ion and fluid scales couple to each other. More than four simultaneous points of measurement are needed to resolve such a coupling. Seven measurement points would be the optimal configuration since they allow to fully resolve scale coupling in 3D. With five or six measurement points, much valuable information can still be obtained by resolving scale coupling in 2D. Another limitation of 4-point constellations is that they can only determine the motion/orientation of structures in 3D at a single scale under the assumptions of planarity and stationarity (Paschmann 1998; Paschmann 2008). More than four measurements are needed to resolve nonplanarity and nonstationarity in 3D.

The need for new multi-scale constellations having more than 4 points is recognized worldwide. Different multi-scale mission concepts are included in the roadmaps of several international space agencies. The NASA [Helioswarm](Helioswarm) mission has been recently selected for being launched in 2028. Helioswarm has an orbit and instrumentation tailored to study turbulence in the solar wind and cannot fully address particle energization and energy transport mechanisms in the more complex plasmas of the Magnetospheric System. A constellation such as Plasma Observatory, having a more advanced instrumentation and an orbit tailored to cover the Magnetospheric System, will be the next logical step after Helioswarm.

**Key science questions**

1) **How are particles energized in space plasmas ?**

Studying plasma energization is of key importance for understanding how our planet works, e.g. how the Earth's radiation environment evolves and how charged particles escape from our ionosphere/atmosphere, as well as for understanding harmful impacts on essential technological systems and human health. Particle energization is related to shocks, magnetic reconnection, turbulence and waves, and plasma jets. The exact physical mechanisms of energization behind these fundamental processes, as well as their combination, are not yet understood (Retinò 2021).

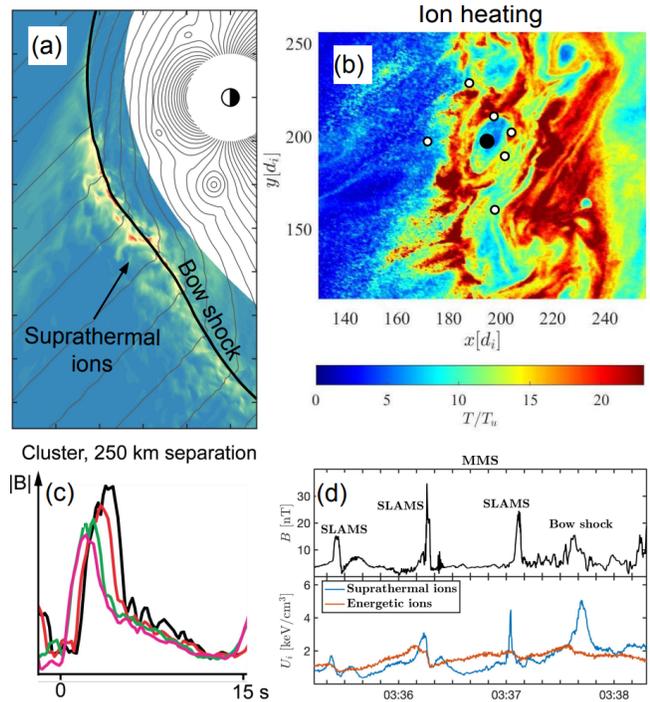

*Figure 2. Ion energization at shocks. Supercomputer kinetic simulations of (a) terrestrial bow shock (Johlander 2021), (b) local fluid/ion scale (Trotta 2021), $d_i$ is the ion scale. Circles illustrate possible Plasma Observatory constellation projected in the equatorial plane. Observations of SLAMS and ion energization as seen in (c) Cluster (Lucek 2008) and (d) MMS data (Johlander 2021).*

The specific science questions addressed by Plasma Observatory are: *How are particles energized (1) at shocks, (2) during magnetic reconnection, (3) by waves and turbulent fluctuations, (4) in plasma jets, and (5) upon combination of these processes?* All these questions involve coupling across multiple scales from system-size down to electron scales. However, fluid and ion scales are the scales at which the largest amount of electromagnetic energy is converted into particle energization. In addition, plasma structures important for energization, e.g., filaments, vortices and current sheets, are most of the time nonplanar and nonstationary. Resolving complex scale coupling, as well as nonplanarity and nonstationarity, is required to reach closure on these questions. This cannot be achieved with the current Cluster, THEMIS and MMS constellations.

Figure 2 shows a key open example: ion injection to suprathermal energy at the bow shock. One example of efficient ion injection can be provided by Short Large Amplitude Magnetic Structures (SLAMS), see Figure 2 d, which are nonplanar and nonstationary shock structures in between ion and fluid scales (Johlander 2016; Schwartz 1991). Cluster 4-point measurements at ion scales show that SLAMS are not consistent with a planar and stationary structure which is simply propagating across the spacecraft, implying a more complex spatiotemporal dynamics (Figure 2c). MMS has much better time resolution of ion measurements and resolves ion energization, however, the MMS inter-spacecraft separation is too small to study SLAMS dynamics. More than 4 points are required to resolve the nonplanarity and nonstationarity of SLAMS.

2) **Which processes dominate energy transport and drive coupling between the different regions of the Earth's magnetospheric system?**

The solar-wind energy enters continuously, at a variable rate, the boundaries of the magnetosphere, is then injected in its inner part, and eventually couples with the ionosphere/atmosphere. Understanding energy transport processes is therefore of key importance for ultimately comprehending the dynamic of the near-Earth's plasma environment. This is also of value since similar processes occur in planetary magnetospheres (Kotsiaros 2019; Slavin 2010) and likely also in other planetary systems (Ben-Jaffel 2022). The energy transport and coupling between different magnetospheric regions is realized through mechanisms which comprise plasma jets, Field-Aligned Currents (FACs), and plasma instabilities and which are still not understood.

The specific science questions addressed by Plasma Observatory are: *(1) How do plasma jets interact with the Earth's dipole field in the transition region? (2) How do field-aligned currents connect different regions of the Magnetospheric System? (3) Which are the key plasma instabilities involved in energy transport? (4) How is energy flux partitioned in different energy transport processes?*

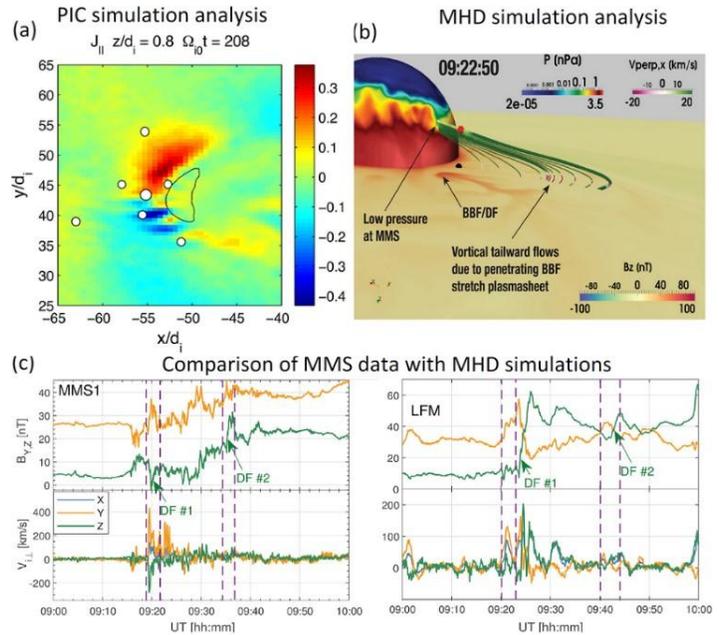

*Figure 3. Plasma jet interaction with the Earth's dipolar field in the transition region: (a) kinetic simulations showing coupling between ion and fluid scales during field-aligned current generation at jet heads (Pritchett 2017); Plasma Observatory possible constellation superimposed. MHD simulations (b) and MMS observations (c, left) compared with virtual spacecraft observations in (b) and (c, right) showing complex nonplanar/nonstationary BBFs/DFs structure (Merkin 2018).*

The transition region is a key example where energy transport and plasma coupling give rise to major effects in the magnetosphere-ionosphere system. In this region, the high-speed plasma jets (Angelopoulos 2013; Baumjohann 1990; Khotyaintsev 2011), often referred to as Bursty Bulk Flows (BBFs), brake (Shiokawa 1997) and/or divert (Juusola 2011), possibly generating FACs. Major particle acceleration occurs in this region (Ukhorskiy 2017) and simulations suggest that the transition region determines the particle injection in the ionosphere (Grandin 2019). Also, different auroral forms are linked to the transition region (Forsyth 2020). In situ data and simulations suggest that many key physical processes occurring in the transition region in the presence of BBFs involve coupling between ion and fluid scales such as generation of FACs, see also Figure 3a (Birn 2020; Nakamura 2021). Many open questions are still related to all these processes and resolving coupling between ion and fluid scales is essential. BBFs propagation also generates dipolarization fronts (DFs) that are often 3D nonplanar and nonstationary, see Figure 3c. Fluid-scale Cluster observations (Nakamura 2004) have shown that BBFs/DFs may reach a few Earth's radii extension in the cross-tail direction. These observations, coupled with the high-time resolution MMS measurements, provided evidence that ion and sub-ion scales processes are important in forming DFs (Fu 2012). Yet, these previous four-point observations can study either the substructure of the DFs or the fluid-scale structure of BBFs, depending on the separation of the spacecraft. For complete understanding of the processes in the transition region, both the fluid-scale structure of BBF/DFs and the ion/sub-ion scale DF processes need to be observed simultaneously and more than four measurement points are needed.

**The Plasma Observatory mission concept**

The Plasma Observatory constellation comprises one mothercraft (MSC) and six to four identical smallsat daughtercraft (DSC) (Marcucci 2022). The science orbit is HEO 8 RE ×18 RE with an inclination of 15° with respect to the equatorial plane. This science orbit permits to maximize the possibilities of making measurements in the Key Science Regions (KSRs) of the Magnetospheric System: the foreshock, bow shock, magnetosheath, magnetopause, magnetotail current sheet, and transition region. Plasma Observatory constellation permits simultaneous observations at the ion and fluid scales. The nominal mission duration is 3 years.

The Plasma Observatory spacecraft will be equipped with state-of-the art instrumentation for the measurement of the electric and magnetic fields and particle distribution functions. The MSC payload comprises fluxgate and search coil magnetometers, an electric field instrument and particle spectrometers for the measurement of electrons, mass resolved ions and energetic particles. The MSC payload provides high time resolution measurements enabling the study of sub-ion scales. The DSC payload gives a complete description of the physical parameters at ion and fluid scales. It is composed of a fluxgate magnetometer, an electric field instrument, an ion and electron spectrometer, and an energetic particle instrument. The Plasma Observatory MSC is a spin stabilized spacecraft spinning at 2 rpm, with a Sun-oriented spin axis, while the DSCs are spin stabilized spacecraft spinning at 15 rpm, with a spin axis perpendicular to the ecliptic.

**International collaborations**

International collaborations are essential to build a network of satellites simultaneously covering scales from mesoscales to ion kinetic scales in the Earth's Magnetospheric System, as highlighted within the COSPAR Task Group on Establishing an International Geospace Systems Program (TGIGSP) aiming at such possibility for mid/late 2030s to realize the International Solar Terrestrial Program Next (ISTPNext) (Kepko et al. 2022a).

MagCon (Magnetospheric Constellation) is a mission concept being studied by NASA for a possible launch in mid/late 2030s (Kepko et al 2022b). MagCon aims at studying the flow of mass, momentum, and energy through the Earth's magnetosphere at mesoscales (1-3 RE). The baseline configuration consists of 36 identical spacecraft placed in the equatorial plane with a spacecraft separation of ~1 RE across a +/- 5 RE width near apogee. Observations at mesoscales by MagCon are strongly complementary to those at the ion and fluid scales which could be performed by the Plasma Observatory.

Further synergies may come from collaboration with Japan and China. The Japanese NEO-SCOPE constellation, which also might be launched by mid/late 2030s. NEO-SCOPE is dedicated to study scale coupling in reconnection and shocks and is currently being discussed for the roadmap of the Japanese Solar Terrestrial Physics Group as a possible JAXA middle class mission. The Chinese AME mission is currently being studied for a possible launch in early 2030s (Lei 2020). AME is a constellation of 1 mothercraft and 12 cubesat daughtercraft aiming to study scale coupling in magnetic reconnection.

Coordination between Plasma Observatory and MagCon mission, together with possible additional synergies with NEO-SCOPE, AME and other multi-scale constellation concepts, would allow us, for the first time, to simultaneously cover the physics of plasma energization and energy transport in the Earth's Magnetospheric System from the large mesoscales to small ion kinetic scales, leading to a paradigm shift in the field of space plasma and magnetospheric physics.